\documentclass[11pt,preprint]{aastex}
\usepackage{graphics,lscape}
\usepackage{color}
\usepackage{natbib}
\usepackage{fancybox}
{\begin{Sbox}\begin{minipage}}%
{\end{minipage}\end{Sbox}\fbox{\TheSbox}}
\usepackage{tikz}
\usetikzlibrary{shapes}
\usepackage{epsfig}
\usepackage{slashbox}
\usepackage{multirow}
\usepackage{lscape}
\usepackage{mathrsfs,amssymb}
\usepackage{amsmath}
\usepackage{datetime}

\usepackage{marginnote}
\let\oldmarginpar\marginpar
\renewcommand\marginpar[1]{\-\oldmarginpar[\raggedleft\scriptsize #1]%
{\raggedright\scriptsize\color{red} #1 \\\rule[4pt]{60pt}{1pt}}}

\slugcomment{v2.3 \today~\currenttime}

\newcommand{\spitzerirs}{{\em Spitzer}/IRS\ }

\newcommand{\etal}{\textrm{et al. }}

\newcommand{\eg}{\textrm{e.g. }}

\newcommand{\um}        {\,\mu{\rm m}}
\newcommand{\mum}        {\,\mu{\rm m}}

\newcommand{\pc}        {\,{\rm pc}}

\newcommand{\Msun}      {\,{\rm M_\odot}}

\newcommand       \simali       {\,{\sim}}

\newcommand       \K        {\,{\rm K}}

\newcommand       \kabs          {\kappa_{\rm abs}}

\newcommand       \simgt        {\gtrsim}

\begin{document}
\title{A TALE OF THREE GALAXIES: ANOMALOUS DUST PROPERTIES 
IN IRAS\,F10398+1455, IRAS\,F21013-0739 AND SDSS\,J0808+3948}
\author{
Yanxia~Xie\altaffilmark{1,2},
Lei~Hao\altaffilmark{1\dag}, and
Aigen~Li\altaffilmark{2}
}
\altaffiltext{1}{Shanghai Astronomical Observatory, 
                       Chinese Academy of Sciences, 
                       80 Nandan Road, Shanghai 200030, China
                       }
\altaffiltext{2}{Department of Physics and Astronomy, 
                       University of Missouri, 
                       Columbia, MO 65211, USA
                       }
\altaffiltext{\dag}{haol@shao.ac.cn}

\begin{abstract}
On a galactic scale the 9.7$\mum$ silicate {\it emission} 
is usually only seen in type 1 active galactic nuclei (AGNs). 
They usually also display a {\it flat} emission continuum 
at $\simali$5--8$\mum$ 
and the absence of polycyclic aromatic hydrocarbon (PAH) 
emission bands.  In contrast, starburst galaxies, 
luminous infrared (IR) galaxies (LIRGs), 
and ultraluminous IR galaxies (ULIRGs) 
exhibit a {\it red} 5--8$\mum$ emission continuum, 
strong 9.7$\mum$ and 18$\mum$ silicate 
{\it absorption} features, 
and strong PAH emission bands. 
Here we report the detection of anomalous dust properties 
by {\it Spitzer}/Infrared Spectrograph in three galaxies 
(IRAS\,F10398+1455, IRAS\,F21013-0739 and SDSS\,J0808+3948) 
which are characterized by the simultaneous detection of 
a {\it red} 5--8$\mum$ emission continuum, 
the 9.7 and 18$\mum$ silicate {\it emission} features 
as well as strong PAH emission bands. 
These apparently contradictory dust IR emission properties
are discussed in terms of iron-poor silicate composition, 
carbon dust deficit, small grain size and low dust temperature 
in the young AGN phase of these three galaxies.
\end{abstract}



\section{Introduction\label{intro}}                                                
Active galactic nuclei (AGN) are postulated to 
have their central powering sources blocked by
a dust torus. The dust reprocesses the ultraviolet
(UV) and optical photons from the central engine 
and re-radiates in the infrared (IR). 
Different types of AGNs are proposed to 
be the same type of objects but viewed at different 
angles between the line of sight of an observer and 
the central engine (Antonucci 1993). 

The 9.7$\um$ and 18$\um$ emission features detected 
in type 1 AGNs (Sturm \etal 2005, Siebenmorgen \etal 2005, 
Hao \etal 2005) are believed to originate from warm amorphous 
silicate dust. In AGNs, some dust is hot enough to generate 
a strong, {\it flat} emission continuum 
in the 5--8 $\um$ wavelength range, 
making it an effective diagnostic tool 
to distinguish AGN-dominated galaxies 
from starburst-dominated galaxies 
(e.g., see Laurent \etal 2000, Nardini \etal 2008). 
In addition to silicate dust, polycyclic aromatic hydrocarbon
(PAH) molecules also display rich emission bands 
in the mid-IR (Tielens 2008, Wang et al.\ 2014). 
These  features are usually very weak or absent in AGNs 
because PAHs are believed to have been destroyed 
by the harsh radiation field of AGNs.
Compared to AGNs, starburst galaxies commonly 
display a low, {\it red} 5--8$\um$ emission continuum,
strong PAH emission features 
and silicate {\it absorption} features
(\eg Brandl \etal 2006, Hao \etal 2007, Smith \etal 2007).
 
Here we report the simultaneous detection of 
a low, {\it red} $\simali$5--8$\um$ emission continuum 
and strong silicate {\it emission} features in three galaxies, 
IRAS F10398+1455, IRAS F21013-0739 and SDSS J0808+3948. 
They also exhibit strong PAH emission in the mid-IR. 
The IR spectral characteristics of these galaxies are unique 
or anomalous in the sense that they show silicate {\it emission} 
which is characteristic of AGNs 
while they also show a low, {\it red} $\simali$5--8$\um$ 
emission continuum and strong PAH emission 
which are characteristic of starburst galaxies.
%
%
We summarize the observations and data reduction in \S2.
The analysis and results are presented in \S3.
We discuss the implications of these results in \S4. 
The major conclusions are summarized in \S5.

\section{Observations and Data Reduction}

\begin{deluxetable}{lcccccc}   
\tablecolumns{7}
\tablewidth{0pc}
\tablecaption{Basic Parameters for the Three Sources}
\tablehead{

\colhead{Sources} & \colhead{RA} & \colhead{DEC} & \colhead{Redshift} & 
\colhead{AORKEY} & \colhead{$d_{\rm L}$\tablenotemark{a}} & 
\colhead{$L_{\rm UV}$\tablenotemark{b}}                          \\

\colhead{} & \colhead{} & \colhead{} & \colhead{} & 
\colhead{}  & \colhead{(Mpc)} & \colhead{($10^{43} \rm ergs\,s^{-1})$} 
}
\startdata
IRAS F10398+1455 & 10h42m33.32s & +14d39m54.1s  & 0.099 & 22132992 & 456 & 2.05 \\
IRAS F21013-0739 & 21h03m58.75s & -07d28m02.5s  & 0.136 & 23017216 & 640 & 16.5 \\
SDSS J0808+3948  & 08h08m44.27s & 39d48m52.36s  & 0.091 & 23014144 & 416 & 8.50 \\
\enddata

\tablenotetext{a}{ ~Luminosity distance. }
\tablenotetext{b}{ ~GALEX Near-Ultraviolet (NUV) luminosity 
 (taken from NED).}
\end{deluxetable}
The mid-IR spectra of the three sources were obtained 
with the {\it Infrared Spectrograph} (IRS) instrument on board 
the {\it Spitzer Space Telescope} (Houck \etal 2004). 
The IRS spectra cover the 5--40 $\um$ wavelength range, 
with a spectral resolution that varies from $\simali$57 to 
$\simali$128. We use the data archived in 
the {\it Cornell AtlaS of Spitzer/IRS Sources} (CASSIS) 
where the spectra are extracted according to 
the extent of sources (Lebouteiller \etal 2011). 
The CASSIS atlas includes $\simali$13,000 low resolution 
spectra of $>$11,000 distinct sources observed 
in the standard staring mode 
and provides publishable quality spectra. 
Table~1 presents the basic parameters 
for the three sources.

\section{Results}
We show the IRS spectra of the three sources in Figure~1a. 
They all exhibit strong PAH emission features 
at 6.2$\um$, 7.7$\um$, 8.6$\um$, 11.2$\um$ and 12.7$\um$
as well as the 9.65$\mum$ H$_2$S(3) rotational line.
These are typical emission features of starburst galaxies. 
Several atomic emission features 
which are related to starburst activity 
are also detected in these three sources: 
[Ar II] 6.986$\mum$,
[Ne II] 12.81$\mum$ 
and [Ne III] 15.56$\um$. 
The 10.51$\mum$ [S IV] line is prominent in IRAS F10398+1455 
and IRAS F21013-0739, but not seen in SDSS J0808+3948.
No obvious high ionization-potential lines 
like [Ne V] 14.32$\um$ and [O IV] 25.89$\um$ 
(which are typical in AGNs)
are detected in our three galaxies. 
%
Finally, the 9.7$\mum$ and 18$\mum$ amorphous silicate 
emission features are clearly seen in all three galaxies.

\begin{figure} 
\begin{center}
\resizebox{\hsize}{!}{\includegraphics{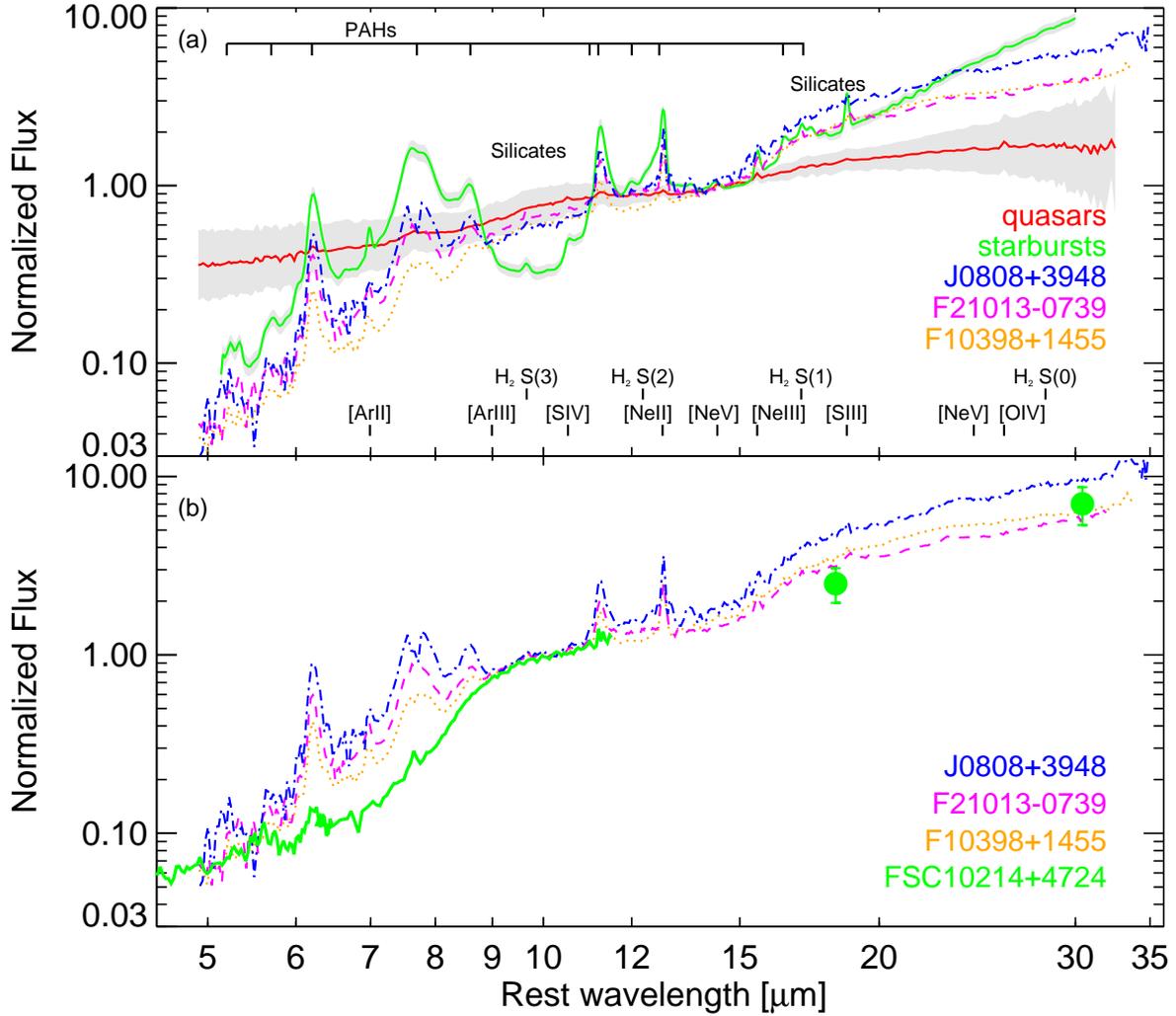}}
\caption{\footnotesize
             Upper panel (a): Comparison of the IRS spectra of 
             SDSS J0808+3948 (blue), IRAS F21013-0739 (magenta), 
             and IRAS F10398+1455 (orange),
             with the average spectrum of quasars 
            (red; Hao \etal 2007) and the average spectrum 
            of starburst galaxies (green; Brandl \etal 2006). 
            All spectra are normalized at 14.5$\mum$.
            Bottom panel (b): 
              Comparison of the IRS spectra of SDSS J0808+3948 (blue),
              IRAS F21013-0739 (magenta), and IRAS F10398+1455 (orange)
              with that of IRAS FSC 10214+4724 
             (green; Teplitz et al. 2006),
              a lensed ULIRG. Also shown are the IRAS 60$\um$ 
              and 100$\um$ photometry of IRAS FSC 10214+4724
              (filled green circles) with its redshift 
              of $z\approx 2.29$ taken into account.
              All spectra are normalized at 10$\mum$.
            }
\end{center}
\end{figure} 

In Figure~1a we also compare the mid-IR spectra of 
these three galaxies with the averaged spectrum of 
starburst galaxies (Brandl \etal 2006) and the average 
spectrum of quasars (Hao \etal 2007). 
It is seen that our three sources have a comparable silicate 
emission strength as quasars, but they show a much steeper 
or ``redder'' 5--8$\mum$ continuum than typical AGNs or quasars.
The slope of the 5--8$\mum$ continuum,
defined as $d\ln F_\nu/d\ln\lambda$
(where $F_\nu$ is the observed flux 
and $\lambda$ is wavelength),
is $\simali$4.1--4.6 for the three sources, 
while the 5--8$\mum$ continuum is 
flat or ``gray'' for AGNs (with a slope of $\simali$0.8). 
The ``red'' 5--8$\mum$ emission continuum and 
the strong PAH emission features seen in 
these galaxies are also seen in starburst galaxies.
However, the 9.7$\mum$ and 18$\mum$ emission features 
seen in these galaxies are never seen 
in starburst galaxies.
Type 1 AGNs (and some type 2 AGNs; 
see Sturm et al.\ 2006, 
Mason et al.\ 2009, 
Nikutta et al.\ 2009, 
Shi et al.\ 2010) 
exhibit the silicate emission features
seen in these galaxies, 
but their 5--8$\mum$ emission continuum 
is often much flatter 
and do not show the PAH emission features
which are prominent in these three galaxies.

The exact profiles and strengths of 
the derived silicate emission features 
are sensitive to the assumed underlying
continuum (e.g., see Sirocky et al.\ 2008,
Baum et al.\ 2010, Gallimore et al.\ 2010).
To subtract the continuum,  
we take two approaches:
(1) we select five points at 5--7$\mum$, 
14.5--15.0$\mum$ and 29--30$\um$ 
to define a underlying continuum 
which is calculated with a spline function
(see Figure~2);
(2) we modify the PAHFIT software of 
Smith et al.\ (2007) to fit the
observed IRS spectrum of each source
with a combination of PAH features,
amorphous silicate features,
modified blackbodies, and starlight,
with the sum of the modified blackbodies
and starlight representing the continuum
underneath the PAH and silicate features
(see Figure~3).
Not surprisingly, the silicate emission 
features derived from these two approaches 
differ substantially: the 9.7$\mum$ feature
derived from the spline method 
is considerably weaker than that derived
from the PAHFIT method, while it is in
the opposite for the 18$\mum$ feature. 
Nevertheless, it is clear that both the 9.7$\mum$
and 18$\mum$ amorphous silicate emission features
are present in these three sources.

Crystalline forsterite silicates are also present 
in these galaxies as revealed by the sharp features 
at $\simali$19, 23, and 27.5$\mum$ (see Figures~2, 3). 
The detection of crystalline silicate dust
in extragalactic sources has been reported 
for some ULIRGs (Spoon et al.\ 2006),
for PG 2112+059, 
a broad absorption line quasar (Kemper et al.\ 2007), 
for a distant absorber at redshift $z_{\rm abs}\approx 0.886$ 
along the line of sight toward the gravitationally lensed quasar 
PKS 1830-211 (Aller et al.\ 2012),
and for a $z_{\rm abs}\approx 0.685$ absorber
toward the gravitationally lensed blazar TXS 0218+357
(Aller et al.\ 2014).

In the Galactic ISM the silicate dust
is predominantly amorphous
(Li \& Draine 2001, Kemper et al.\ 2004, Li et al.\ 2007).
As shown in Figures~2, 3, 
the silicate features of these three galaxies
are broad and smooth 
with sharp crystalline silicate features superimposed,
suggesting that the silicate dust in these sources
is mainly amorphous. 
%


\begin{figure*}  
\begin{center}
\resizebox{\hsize}{!}{\includegraphics{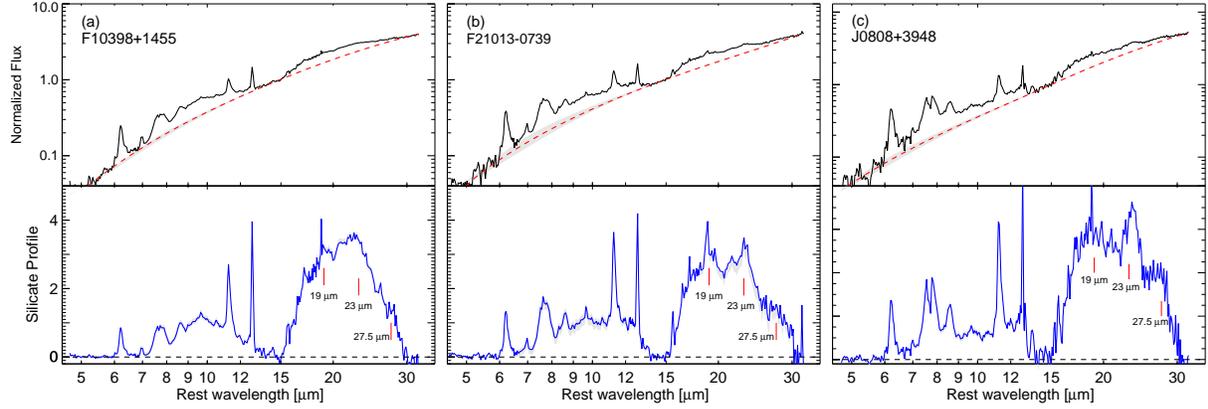}}
\caption{\footnotesize
              The IRS spectra and the underlying 
              {\it spline} continuum 
              (upper panel) as well as the continuum-subtracted 
              silicate emission profiles (bottom panel)
              of IRAS F10398+1455 (a), IRAS F21013-0739 (b)
              and SDSS J0808+3948 (c).
              The PAH features at 6.2, 7.7, 8.6, 11.3 and 12.7$\mum$
              and the crystalline silicate features at 19, 23 and
              27.5$\mum$ are prominent.
              }
\end{center}
\end{figure*}

\begin{figure*}  
\begin{center}
\resizebox{\hsize}{!}{\includegraphics{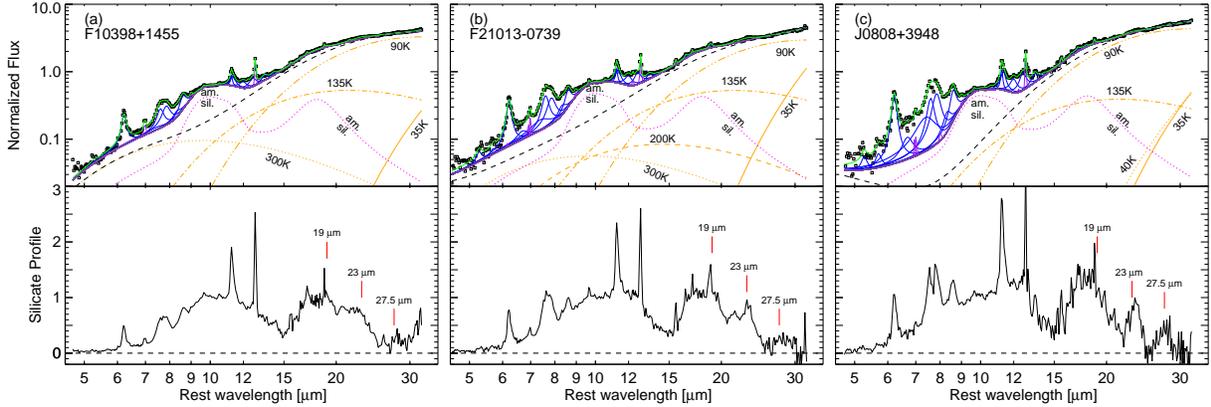}}
\caption{\footnotesize
              Same as Figure~2 
              but with the underlying continuum 
              determined by fitting the IRS spectra 
              with a combination of 
              PAH features (solid blue lines),
              amorphous silicate features
              (dotted magenta lines),
              modified blackbodies of
              different temperatures,
              and starlight (upper panel).
              The sum of the modified blackbodies
              and starlight represents the continuum
              underneath the PAH and silicate features
              (dashed black line).
              The continuum-subtracted 
              silicate emission profiles 
              are shown in the bottom panel.
              }
\end{center}
\end{figure*}

\section{Discussion}
%
%
%
These three galaxies are found when cross-matching 
the SDSS and {\it Spitzer}/IRS low resolution spectra 
(L.~Hao \etal 2014, in preparation).
Xie \etal (2014a) have studied the multi-wavelength properties 
of these three galaxies and demonstrated that they may harbor 
a young AGN at the center. Among these galaxies, 
IRAS F21013-0739 and SDSS J0808+3948 are 
{\it Lyman Break Analogs} 
(LBA; Heckman \etal 2005, Hoopes \etal 2007). 
LBAs are rare in the local universe.
They are selected based on their far-UV properties 
and they show similar properties as that of 
high-redshift {\it Lyman Break Galaxies} 
(Steidel et al.\ 1999). 
Among the 30 LBA galaxies, six are found to 
contain a young, very compact ($\simali$10$^2\pc$), 
highly massive (several 10$^9\Msun$) 
{\it Dominant Central Object} (DCO, Overzier \etal 2009). 
Interestingly, both F21013-0739 and SDSS J0808+3948 
are found to have DCOs. It appears that the occurrence of
the anomalous mid-IR spectral character is associated 
with the galaxies with a DCO. 
IRAS F10398+1455 is not a LBA. 
This galaxy is very faint in the GALEX Far-UV (FUV) band 
but it is bright in the GALEX NUV band. We list the NUV 
luminosity in Table 1 for the three galaxies. 
It is unresolved in the NUV image and its light distribution 
is comparable to that of IRAS F21013-0739 and 
that of the GALEX standard star, 
indicating a compact UV region in this galaxy. 

The type of spectrum  
(i.e., a combination of 
a {\it red} 5--8$\mum$ continuum
with strong silicate {\it emission})
found for the three galaxies 
reported here is very rare.
To the best of our knowledge, the only similar source 
is IRAS FSC 10214+4724, a lensed ULIRG at
a redshift of $z\approx 2.29$ (Teplitz \etal 2006),
which shows a red 5--8$\um$ emission continuum 
and strong silicate emission features. 
However, little or no PAH emission is seen in this source
(see Figure~1b).
Teplitz \etal (2006) showed that the silicate emission 
features of this source are similar to that of other AGNs 
observed by {\it Spitzer}/IRS.
They fitted the near-IR to $\lambda\approx365\mum$
rest-frame spectral energy distribution (SED) 
of this source with a combination of 
hot dust ($T\approx 640\K$), 
warm dust ($T\approx 190\K$) 
and cold dust ($T\approx50\K$). 
They found that the contribution of
the hot dust component to the SED 
which is directly related to AGN heating 
and accounts for the 2--8$\mum$ continuum 
emission is rather small compared to 
other AGN-dominated ULIRGs and QSOs. 
%
%
If most of the IR luminosity of IRAS FSC 10214+4724
is due to starburst, it is difficult to understand 
why the PAH emission which is strong in starbust galaxies 
is rather weak in FSC 10214+4724. 

Teplitz \etal (2006) explained 
the unusual spectral characteristics 
of IRAS FSC 10214+4724 
(i.e., a {\it red} 2--8$\mum$ continuum, 
strong silicate {\it emission},
little or no PAH emission)
in terms of differential lens magnification
that magnifies the central AGN features
(e.g., the silicate emission), 
while masking the starburst signatures 
with unusually weak PAH emission. 
But this hypothesis could not explain
the weak, red 2--8$\mum$ continuum 
(i.e., the hot dust component)
which is believed to be closely tied with 
the central AGN. 

The differential lens magnification scenario 
may not work for our three sources
since all three galaxies are at low redshifts, 
from $z\approx0.091$ to $z\approx0.137$, 
and there are no galaxy clusters lying 
along the line of sight toward these three galaxies 
(see Yang \etal 2007).

\begin{figure}[ht]  
\begin{center}
$
\begin{array}{cc} 
\resizebox{0.5\hsize}{!}{\includegraphics{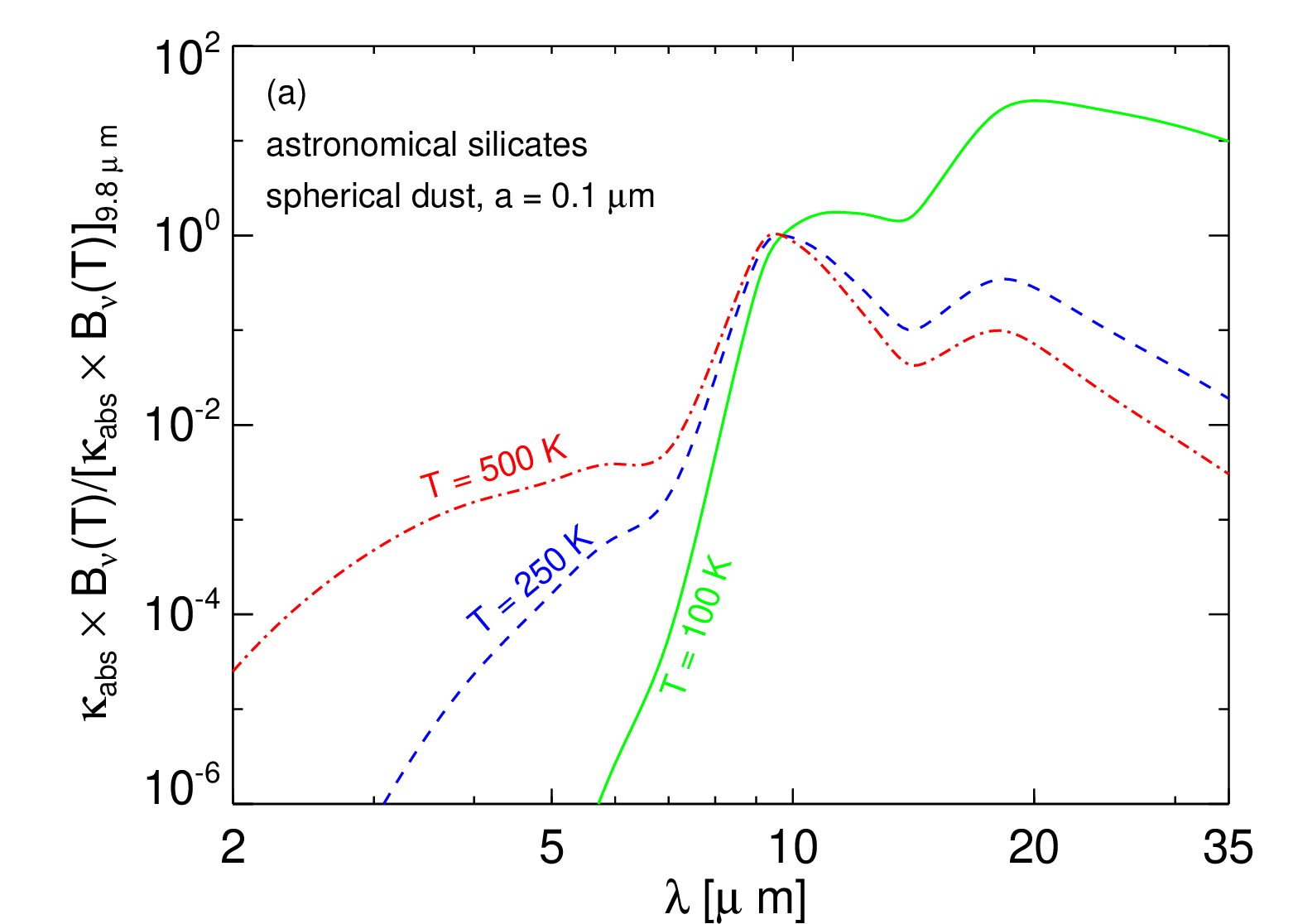}} 
& \resizebox{0.5\hsize}{!}{\includegraphics{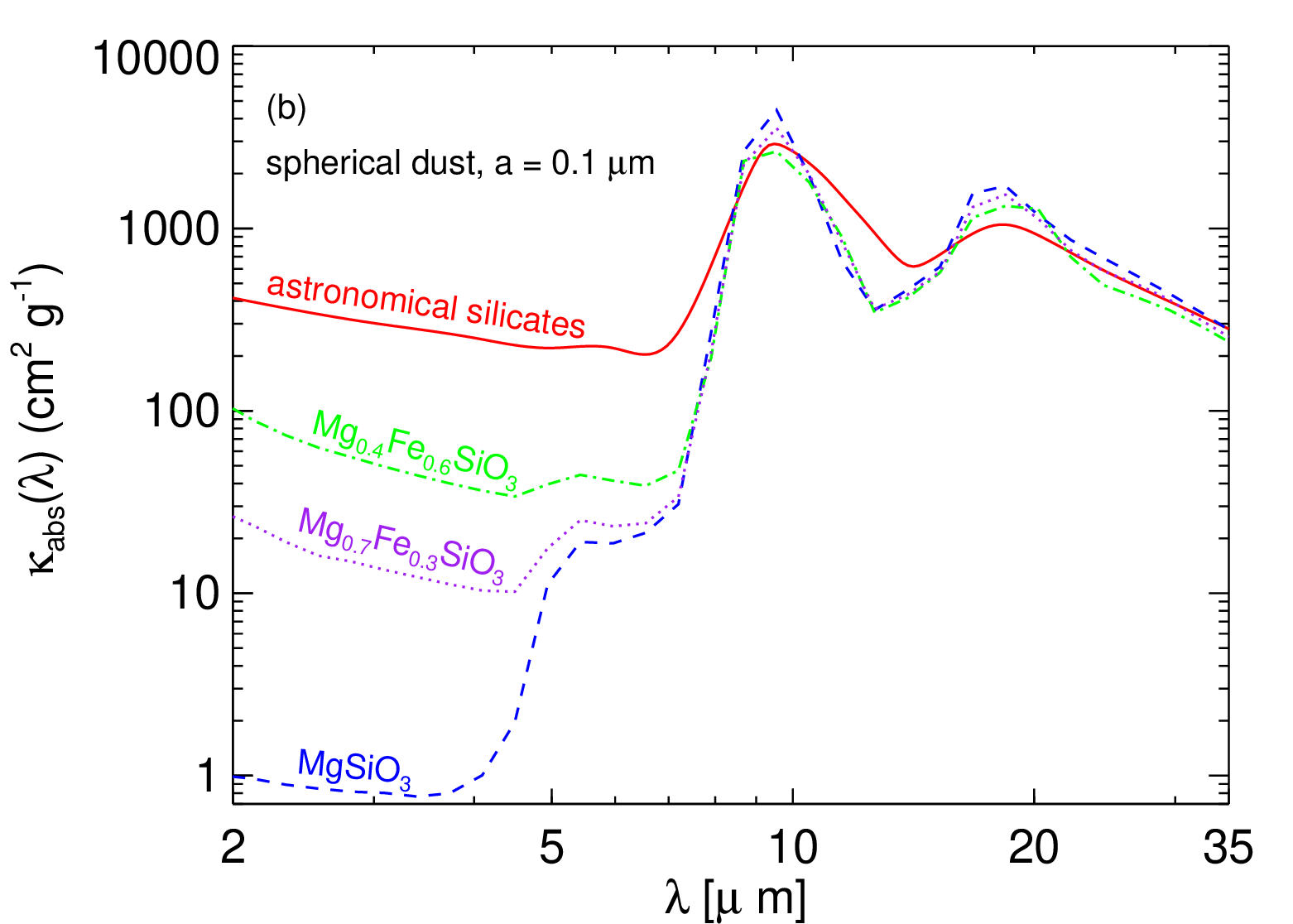}} \\
\resizebox{0.5\hsize}{!}{\includegraphics{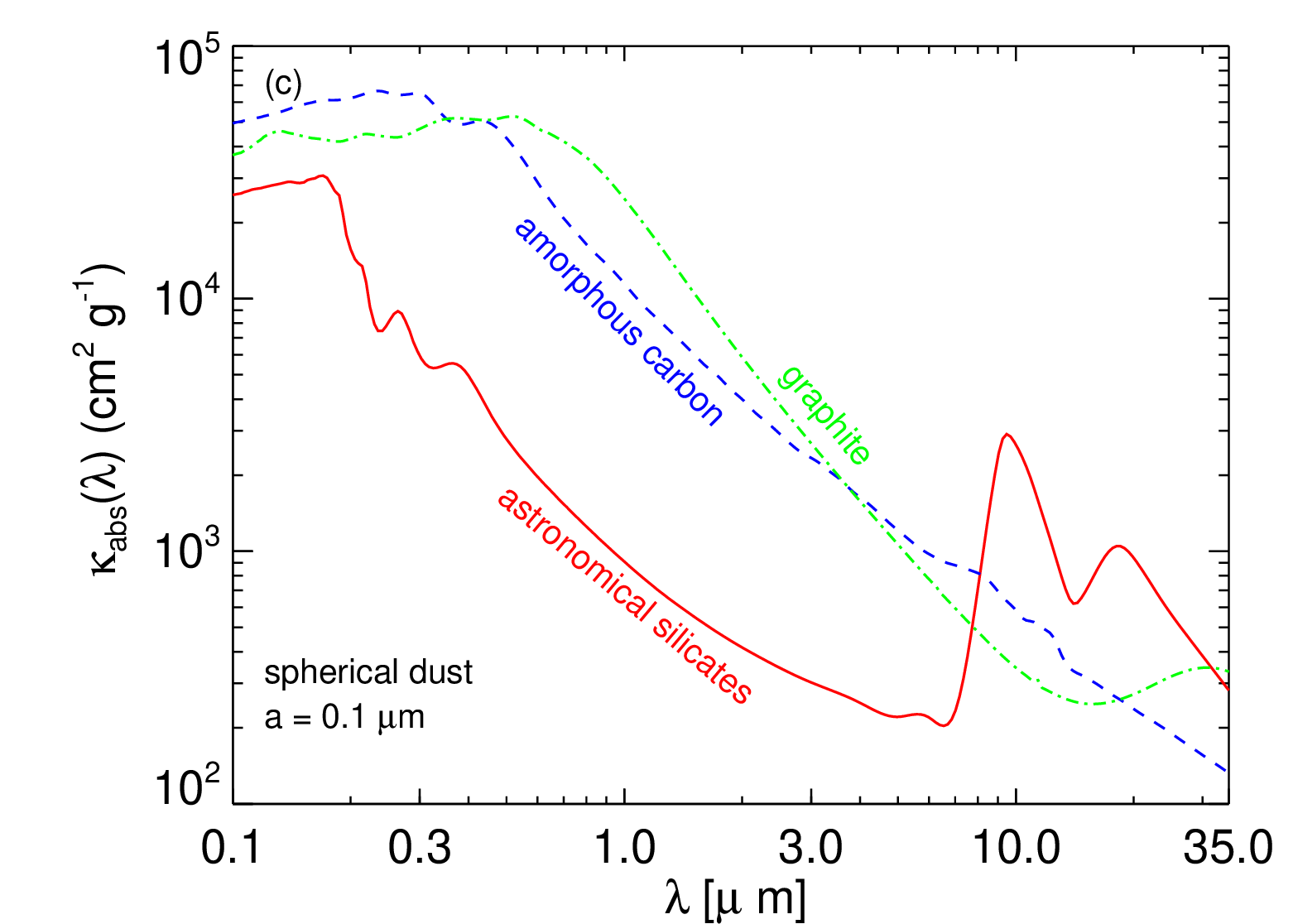}} 
& \resizebox{0.5\hsize}{!}{\includegraphics{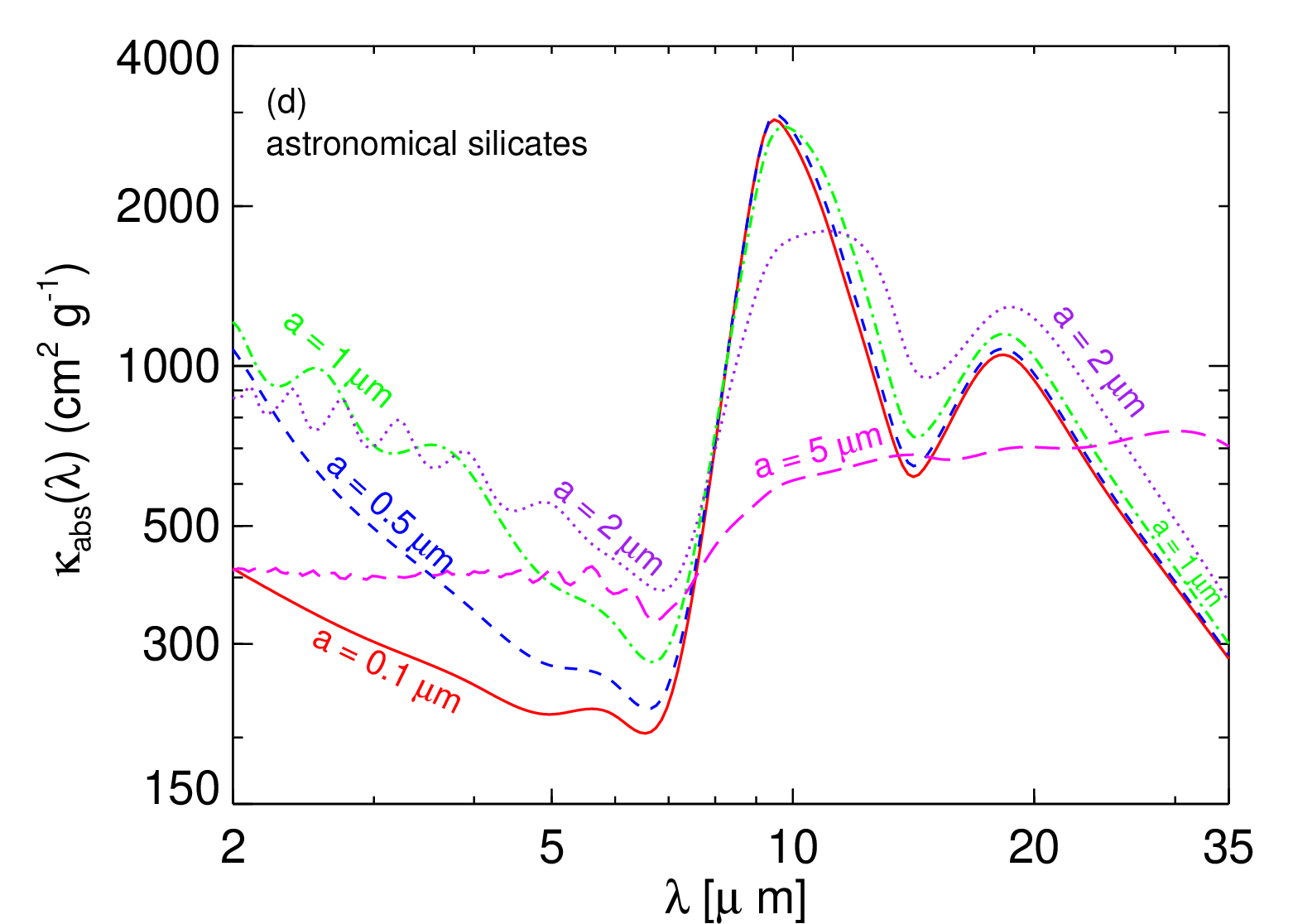}} 
\end{array}
$
\caption{\footnotesize
             Upper left panel (a): Normalized IR emission
             spectra of spherical ``astronomical silicates''
             of radius $a=0.1\mum$ of temperatures
             $T$\,=\,100$\K$ (green solid), 250$\K$ (blue dashed), 
             and 500$\K$ (red dot-dashed).
             Upper right panel (b): Mass absorption coefficients
             of ``astronomical silicates'' (red solid)
             and amorphous pyroxenes 
             of varying iron contents:
             MgSiO$_3$ (blue dashed) , 
             Mg$_{0.7}$Fe$_{0.3}$SiO$_3$ (purple dotted), and
             Mg$_{0.4}$Fe$_{0.6}$SiO$_3$ (green dot-dashed). 
             All dust species are taken to be spherical with $a=0.1\mum$.
             Bottom left panel (c): Mass absorption coefficients
             of ``astronomical silicates'' (red solid),
             amorphous carbon (blue dashed),
             and graphite (green dot-dashed).
             All dust species are taken to be spherical 
             with $a=0.1\mum$.
             Bottom right panel (d): Mass absorption coefficients
             of spherical ``astronomical silicates'' of various sizes:
             $a=0.1\mum$ (red solid),
             $a=0.5\mum$ (blue dashed),
             $a=1\mum$ (green dot-dashed),
             $a=2\mum$ (purple dotted), and
             $a=5\mum$ (magenta long dashed).
              }
\end{center}
\end{figure}



The combination of
a steeply rising red 5--8$\mum$ emission continuum 
with the silicate emission features at 9.7 and 18$\mum$
suggests the dust temperature is in the range of 
$\simali$100--300$\K$. 
In contrast, the flat 5--8$\mum$ emission and 
the silicate emission of quasars shown in Figure~1a 
imply a dust temperature of $\simali$300--500$\K$.
For illustration, we show in Figure~3a 
the thermal IR emission of 
spherical ``astronomical silicates'' 
(Draine \& Lee 1984) of radius $a=0.1\mum$
at temperatures $T$\,=\,100, 250, and 500$\K$.
The dust IR emission is expressed as 
$\kabs(\nu)\times B_\nu(T)$ and normalized to 
that at $\lambda=9.8\mum$, 
where $\kabs(\nu)$ is the mass absorption coefficient
of the dust at frequency $\nu=c/\lambda$ 
($c$ is the speed of light), 
and $B_\nu(T)$ is the Planck function
of temperature $T$ at frequency $\nu$.
Figure~3a shows that with the increase of $T$,
the 2--8$\mum$ continuum emission becomes 
stronger and flatter. 

The temperature of the dust in an AGN torus 
ranges from the sublimation temperature of
$T\approx 1500\K$ in the inner wall to
$T\approx 100\K$ in the outer boundary (Li 2007).
Depending on how the dust is spatially distributed
in the torus, the appearances of the 5--8$\mum$ continuum
and the silicate emission features would differ from one
to another. 
For the three galaxies reported here,
a simple explanation would be that the dust tori 
of the AGNs in these galaxies 
are at a larger distance from
the central engine compared 
to that of typical AGNs and quasars.
%


Alternatively, the difference between the average 
IRS spectrum of quasars and the IRS spectra of 
our three galaxies could be due to different silicate
composition. Dorschner \etal (1995) have experimentally
demonstrated that the absorption in the UV, optical 
and near-IR up to $\lambda<8\mum$ of amorphous 
silicate dust decreases with the decrease of the iron 
content in the dust.
As shown in Figure~4 of Dorschner \etal (1995), 
the imaginary part of the index of refraction of 
${\rm MgSiO_{3}}$ is smaller than that of
${\rm Mg_{0.4}Fe_{0.6}SiO_{3}}$ by a factor of $>10$.
It is possible that the silicate dust in quasars 
and typical AGNs are iron-rich, while the silicate
dust in the three galaxies are iron-poor.
In Figure~3b we show the mass absorption coefficients
calculated for three amorphous pyroxene dust species:
MgSiO$_3$, Mg$_{0.7}$Fe$_{0.3}$SiO$_3$, and
Mg$_{0.4}$Fe$_{0.6}$SiO$_3$, 
with the refractive indices 
taken from Dorschner et al.\ (1995).
All dust species are assumed to be spherical 
and have a radius of $a=0.1\mum$.
It is apparent that the continuum absorption at 
$\lambda<8\mum$ substantially increases with
the iron content. We note that the Draine \& Lee (1984)
``astronomical silicates'' were synthesized to be ``dirty''
so that they are absorptive in the UV, optical and near-IR.

The 5--8$\mum$ emission continuum difference
between these three galaxies and typical AGNs
and quasars may also be accounted for 
by different fractions of amorphous carbon 
or graphite dust.
In addition to amorphous silicate dust, 
there must be a population of carbonaceous dust
in the Galactic ISM (see Li 2005). 
Although the exact nature of
the carbonaceous dust species is unknown, 
graphite and amorphous carbon are commonly 
suggested as candidates.  
In the 0.1--8$\mum$ wavelength range,
as shown in Figure~3c,
both amorphous carbon and graphite have
a much higher absorption opacity than 
amorphous silicate dust 
(including Fe$_2$SiO$_4$ and FeSiO$_3$).
We take the indices of refraction of
amorphous carbon and graphite
respectively from Rouleau \& Martin (1991)
and Draine \& Lee (1984).
The low 5--8$\mum$ continuum emission
observed in these three galaxies 
could be attributed to a relative deficit of
carbon dust compared to that of typical AGNs/quasars.

The IRS spectral appearance difference between 
these three galaxies and typical AGNs/quasars
could also be due to dust size effects. 
Dust absorbs and scatters light most effectively
at wavelengths comparable to its size 
($2\pi a/\lambda\sim 1$; see Bohren \& Huffman 1983).
As illustrated in Figure~3d, the absorption 
of large ``astronomical silicate'' dust of 
$a\simgt 1\mum$ in the 2--8$\mum$ wavelength range
substantially exceeds that of submicron-sized dust.
For silicate dust of $a\simgt5\mum$,
the absorption at $\lambda<8\mum$ is
``gray'' and the 9.7$\mum$ silicate feature 
is very broad and weak.
The strong, flat 5--8$\mum$ emission continuum 
seen in quasars could be dominated 
by dust of sizes $a\simgt1\mum$,
while the low, red 5--8$\mum$ continuum seen in
these three galaxies may indicate that they are
rich in submicron-sized dust. 



We note that the above analysis on the dust composition
is somewhat simplified. 
It is known that radiative transfer effects 
could account for some of the 
``anomalous'' spectral characteristics of AGNs, 
e.g., Nikutta et al.\ (2009) demonstrated that 
the apparent red-shift of the 
9.7$\mum$ silicate emission feature of AGNs 
(Hao et al.\ 2005, Siebenmorgen et al.\ 2005, 
Sturm et al.\ 2005) 
could be due to radiative transfer effects, 
although it could also be due to porous dust 
(see Li et al.\ 2008) or micrometer-sized dust 
(see Smith et al.\ 2010). 
To more reliably determine the dust composition
from the observed IRS spectrum of a source,
one needs to combine the dust optical properties
with proper radiative transfer. 
However, we have demonstrated 
in Y.~Xie et al.\ (2014, in preparation)
that it is not possible to simultaneously fit
both the silicate emission features 
and the red 5--8$\mum$ continuum 
with the radiative transfer models of 
Nenkova et al.\ (2008a, b) for clumpy tori
and with the silicate optical properties of
Ossenkopf et al.\ (1992).
Models that fit the silicate emission well 
will always overestimate the 5--8$\mum$ 
continuum observed in the three sources.
Therefore, we believe that the anomalous spectral 
characteristics (i.e., a red 5--8$\mum$ continuum, 
strong silicate emission, strong PAH emission) of
these three sources are potentially revealing 
anomalous dust properties intrinsic to them.

Finally, we note that both PAH and silicate emission features 
are strong in the three galaxies (see Figure~2),
in contrast to the absence of PAH emission in
normal AGNs. It is generally believed that PAHs 
are destroyed by the AGN hard radiation field.
The survival of PAHs in these three galaxies 
may be due to the fact that they are at the young 
AGN/starburst phase of their evolution.
Or alternatively, the region covered by 
the {\it Spitzer}/IRS slit may contain parts of 
the starburst region which emits strongly 
in PAH features. 

\section{Summary}\label{sec:summary}
We report {\it Spitzer}/IRS detections of anomalous dust properties 
in three galaxies (IRAS F10398+1455, IRAS F21013-0739,
and SDSS J0808+3948) which harbor young AGN and starburst activity.
 Unlike most other starburts or AGNs observed thus far 
with {\it Spitzer}/IRS, these three galaxies
exhibit a red 5--8$\mum$ emission continuum 
and strong silicate and PAH emission.
The red 5--8$\mum$ emission continuum 
and PAH emission seen in these galaxies
are characteristic of starbursts.
However, for starbursts the silicate features are 
often seen in absorption.
The silicate emission seen in these galaxies
is characteristic of AGNs,
but for AGNs the 5--8$\mum$ emission continuum 
is often flat (or ``gray'') and the PAH emission
features are often absent. 
We argue that these apparently contradictory properties 
of these three galaxies may be explained by
iron-poor silicate dust, a relative deficit of 
amorphous carbon or graphite dust,
and/or cold, submicron-sized dust.

\acknowledgments
We thank J.W.~Lyu, R.~Mason, A.~Mishra 
and the anonymous referee
for helpful suggestions/discussions.
LH and XYX are partially supported by 
the 973 Program of China (2013CB834905, 2009CB824800), 
the Strategic Priority Research Program 
``The Emergence of Cosmological Structures'' 
of Chinese Academy of Sciences (XDB09000000), 
the Shanghai Pujiang Talents Program (10pj1411800) 
and NSFC 11073040. AL and XYX are supported in part 
by NSF AST-1311804 and NASA NNX14AF68G. 
The Cornell Atlas of \spitzerirs Sources (CASSIS) 
is a product of the Infrared Science Center 
at Cornell University, supported by NASA and JPL.



\begin{thebibliography}{}
\expandafter\ifx\csname natexlab\endcsname\relax\def\natexlab#1{#1}\fi

\bibitem[{{Antonucci}(1993)}]{Antoucci1993}
{Antonucci}, R. 1993, \araa, 31, 473

\bibitem[\protect\citeauthoryear{{Aller} et al.}
{{Aller} et al.}{2012}]{aller12}
Aller, M.~C., et al. 2012, \apj, 748, 19

\bibitem[\protect\citeauthoryear{{Aller} et al.}
{{Aller} et al.}{2014}]{aller14}
Aller, M.~C., et al. 2014, \apj, 785, 36

\bibitem[\protect\citeauthoryear{{Baum} et al.}
{{Baum} et al.}{2010}]{baum_etal_10}
Baum, S.~A., et al. 2010, \apj, 710, 289
 

\bibitem[\protect\citeauthoryear{{Brandl} et al.}
{{Brandl} et al.}{2006}]{bra_etal_06}
{Brandl}, B.~R., et al. 2006, \apj, 653, 1129

\bibitem[\protect\citeauthoryear{Bohren \& Huffman}
{Bohren \& Huffman}{1983}]{Bohren&Huffman}
Bohren, C.~F. \& Huffman, D.~R. 1983, 
Absorption and scattering of light by small particles, Wiley

\bibitem[\protect\citeauthoryear{Draine \& Lee}
{Draine \& Lee}]{D&L1984}
Draine, B.~T. \& Lee, H.~M. 1984, \apj, 285, 89


\bibitem[\protect\citeauthoryear{{Dorschner} et al.}
{{Dorschner} et al.}{1995}]{dors_etal_95}
{Dorschner}, J., et al. 1995, \aap, 300, 503

\bibitem[\protect\citeauthoryear{{Gallimore} et al.}
{{Gallimore} et al.}{2010}]{gallimore_etal_10}
Gallimore, J.~F., et al. 2010, \apjs, 187, 172

\bibitem[\protect\citeauthoryear{{Hao} et al.}{{Hao}
  et al.}{2005}]{hao_etal_05}
{Hao}, L., et al. 2005, \apjl, 625, L75

\bibitem[\protect\citeauthoryear{{Hao} et al.}{{Hao}
  et al.}{2007}]{hao_etal_07}
{Hao}, L., et al. 2007, \apjl, 655, L77

\bibitem[\protect\citeauthoryear{{Heckman} et al.}
{{Heckman} et al.}{2005}]{heckman05}
Heckman, T.~M., et al. 2005, \apjl, 619, L35


\bibitem[Hoopes et al.(2007)]{hoopes07}
Hoopes, C., et al. 2007, \apjs, 173, 441

\bibitem[\protect\citeauthoryear{{Houck} et al.}{{Houck}
  et al.}{2004}]{hou_etal_04}
{Houck}, J.~R., et al. 2004, \apjs, 154, 18

\bibitem[\protect\citeauthoryear{{Kemper} et al.}
{{Kemper} et al.}{2004}]{kemper04}
Kemper, F., Vriend, W.~J. \& Tielens, A.~G.~G.~M.
 2004, \apj, 609, 826

\bibitem[\protect\citeauthoryear{{Laurent} et al.}{{Laurent}
et al.}{2000}]{lau_etal_00}
{Laurent}, O., et al. 2000, \aap, 359, 887

\bibitem[\protect\citeauthoryear{{Lebouteiller}et al.}{{Lebouteiller}
et al.}{2011}]{Lebou_etal_2011}
{Lebouteiller}, V., et al. 2011, \apjs, 196, 8

\bibitem[\protect\citeauthoryear{Li}
{Li}{2005}]{LiAG05}
Li, A. 2005, Journal of Physics Conference Series, 6, 229

\bibitem[\protect\citeauthoryear{Li}
{Li}{2007}]{LiAG07}
Li, A. 2007, in The Central Engine of Active Galactic Nuclei (ASP Conf. Ser. 373), ed. L. C. Ho \& J.-M. Wang (San Francisco, CA: ASP), 561

\bibitem[\protect\citeauthoryear{Li \& Draine}
{Li \& Draine}{2001}]{LiDraine01}
Li, A. \& Draine, B.~T. 2001, \apj, 550, L213

\bibitem[\protect\citeauthoryear{Li, Zhao \& Li}
{Li, Zhao \& Li}{2007}]{limoping07}
Li, M.~P., Zhao, G. \& Li, A. 2007, \mnras, 382, L26

\bibitem[\protect\citeauthoryear{Li, Shi \& Li}
{Li, Shi \& Li}{2008}]{limoping08}
Li, M.~P., Shi, Q.~J. \& Li, A. 2008, \mnras, 391, L49

\bibitem[\protect\citeauthoryear{Molster \& Kemper}
{Molster \& Kemper}{2005}]{molster05}
Molster, F. \& Kemper, C. 2005, \ssr, 119, 3

\bibitem[\protect\citeauthoryear{{Kemper} et al.}
{{Kemper} et al.}{2007}]{kemper07}
Markwick-Kemper, F., et al. 2007, \apjl, 668, L107

\bibitem[\protect\citeauthoryear{{Mason} et al.}
{{Mason} et al.}{2009}]{mason09}
Mason, R. E., et al. 2009, \apjl, 693, L136

\bibitem[\protect\citeauthoryear{{Nardini} et al.}
{{Nardini} et al.}{2008}]{nard_etal_08}
Nardini, E., et al. 2008, \mnras, 385, L130

\bibitem[{{Nenkova} {et~al.}(2008{\natexlab{a}})
{Nenkova}, {Sirocky}, {Ivezi{\'c}}, \& 
{Elitzur}}]{Nenkova2008a}
{Nenkova}, M., {Sirocky}, M.~M., {Ivezi{\'c}}, {\v Z}., \& 
{Elitzur}, M. 2008{\natexlab{a}}, ApJ, 685, 147

\bibitem[{{Nenkova} {et~al.}(2008{\natexlab{b}})
{Nenkova}, {Sirocky}, {Nikutta}, {Ivezi{\'c}}, \& 
{Elitzur}}]{Nenkova2008b}
{Nenkova}, M., {Sirocky}, M.~M., {Nikutta}, R., 
{Ivezi{\'c}}, {\v Z}., \& 
{Elitzur}, M. 2008{\natexlab{b}}, ApJ, 685, 160

\bibitem[\protect\citeauthoryear{{Nikutta} et al.}
{{Nikutta} et al.}{2009}]{nikutta_etal_09}
Nikutta, R., et al. 2009, \apj, 707, 1550


\bibitem[Ossenkopf et al.(1992)]{1992A&A...261..567O} 
Ossenkopf, V., Henning, T., \& Mathis, J.~S.\ 1992, A\&A, 261, 567 

\bibitem[\protect\citeauthoryear{{Overzier} et al.}
{{Overzier} et al.}{2009}]{overzier09}
Overzier, R.~A., et al. 2009, \apj, 706, 203

\bibitem[\protect\citeauthoryear{Rouleau \& Martin}
{Rouleau \& Martin}{1991}]{RM1991}
Rouleau, F. \& Martin, P.~G. 1991, \apj, 377 526

\bibitem[\protect\citeauthoryear{{Shi} et al.}
{{Shi} et al.}{2010}]{shi10}
Shi, Y., et al. 2010, \apj, 714, 115

\bibitem[\protect\citeauthoryear{{Sirocky} et al.}
{{Sirocky} et al.}{2008}]{sirocky08}
Sirocky, M.~M, et al. 2008, \apj, 678, 729

\bibitem[\protect\citeauthoryear{{Smith} et al.}
{{Smith} et al.}{2010}]{smith10}
Smith, H.~A., et al. 2010, \apj, 716, 490

\bibitem[\protect\citeauthoryear{{Smith} et al.}
{{Smith} et al.}{2007}]{smith07}
Smith, J.~D.~T., et al. 2007, \apj, 656, 770

\bibitem[\protect\citeauthoryear{{Spoon} et al.}{{Spoon}
 et al.}{2006}]{spo_etal_06}
Spoon, H.~W.~W., et al. 2006, \apj, 638, 759

\bibitem[\protect\citeauthoryear{{Steidel} et al.}
{{Steidel} et al.}{1999}]{steidel09}
Steidel, C.~C., et al. 1999, \apj, 519, 1

\bibitem[\protect\citeauthoryear{{Sturm} et al.}
{{Sturm}et al.}{2005}]{sturm_etal_05}
{Sturm}, E., et al. 2005, \apjl, 629, L21

\bibitem[\protect\citeauthoryear{{Sturm} et al.}
{{Sturm}et al.}{2006}]{sturm_etal_06}
{Sturm}, E., et al. 2006, \apj, 642, 81

\bibitem[\protect\citeauthoryear{Siebenmorgen, Haas,
Kr{\"u}gel \& Schulz}{2005}]{sieben05}
Siebenmorgen, R., et al. 2005, \aap, 436, L5

\bibitem[\protect\citeauthoryear{{Teplitz} et al.}
{{Teplitz} et al.}{2006}]{tepl_etal_06}
Teplitz, H.~I., et al. 2006, \apjl, 638, L1

\bibitem[{{Tielens}(2008)}]{Tielens}
{Tielens}, A.~G.~G.~M. 2008, \araa, 46, 289

\bibitem[Wang et al.(2014)]{2014SSPMA..44..771L} 
Wang, M.L., Li, A., Xiang, F.Y., Tang, Y., \& Zhong, J.X.\ 2014, 
Scientia Sinica -- Physica, Mechanica \& Astronomica, 44, 771 

\bibitem[\protect\citeauthoryear{{Yang} et al.}
{{Yang} et al.}{2007}]{yangxh07}
Yang, X., et al. 2007, \apj, 671, 153

\end{thebibliography}
\end{document}